\documentclass[aps,prl,twocolumn,reprint,superscriptaddress]{revtex4-1}
\usepackage{amsmath}
\usepackage{amssymb}
\usepackage{graphicx}
\usepackage{hyperref}
\usepackage{url}
\usepackage{color,xcolor}
\usepackage{ulem}
\begin{document}

\title{Pressure-driven phase transition from antiferromagnetic semiconductor to nonmagnetic metal in two-leg
ladders $A$Fe$_2$$X$$_3$ ($A$=Ba or K, $X$=S or Se)}
\author{Yang Zhang}
\author{Lingfang Lin}
\author{Jun-Jie Zhang}
\affiliation{School of Physics, Southeast University, Nanjing 211189, China}
\author{Elbio Dagotto}
\affiliation{Department of Physics and Astronomy, University of Tennessee, Knoxville, TN 37996, USA}
\affiliation{Materials Science and Technology Division, Oak Ridge National Laboratory, Oak Ridge, TN 37831, USA}
\author{Shuai Dong}
\email{Corresponding author: sdong@seu.edu.cn}
\affiliation{School of Physics, Southeast University, Nanjing 211189, China}
\date{\today}

\begin{abstract}
The recent discovery of superconductivity in BaFe$_2$S$_3$ [Takahashi {\it et al.}, Nat. Mater. {\bf 14}, 1008 (2015)]
has stimulated considerable interest in 123-type iron chalcogenides. This material is the first reported iron-based
two-leg ladder superconductor, as opposed to the prevailing two-dimensional layered structures of the iron superconductors
family. Once the hydrostatic pressure exceeds $11$ GPa, BaFe$_2$S$_3$ changes from a semiconductor to a superconductor
below $24$~K. Although previous calculations correctly explained its ground state magnetic state and electronic structure,
the pressure induced phase transition was not successfully reproduced. In this work, our first principles calculations
find that with increasing pressure the lattice constants as well as local magnetic moments are gradually suppressed,
followed by a first-order magnetic transition at a critical pressure, with local magnetic moments dropping to zero
suddenly. Our calculations suggests that the self-doping caused by electrons transferred from S to Fe may play a
key role in this transition. The development of a nonmagnetic metallic phase at high pressure may pave the way to
superconductivity. As extensions of this effort, two other 123-type iron chalcogenides, KFe$_2$S$_3$ and KFe$_2$Se$_3$,
have also been investigated. KFe$_2$S$_3$ also displays a first-order transition with increasing pressure,
but KFe$_2$Se$_3$ shows instead a second-order, or weakly first-order, transition. The required pressures for KFe$_2$S$_3$
and KFe$_2$Se$_3$ to quench the magnetism are higher than for BaFe$_2$S$_3$. Further experiments can confirm the predicted
first-order nature of the transition in BaFe$_2$S$_3$ and KFe$_2$S$_3$, as well as the possible
metallic/superconductivity state in other 123-type iron chalcogenides under high pressure.
\end{abstract}

\maketitle

\section{Introduction}
Since the discovery of superconductivity in fluorine doped LaFeAsO \cite{Kamihara:Jacs}, the iron pnictides and chalcogenides have rapidly developed into one of the main branches of research in the field of unconventional superconductors \cite{Lumsden:Jpcm,Dai:Np,Stewart:Rmp}. Almost all previously reported iron-based superconductors have similar crystal structures, involving a slightly distorted two-dimensional Fe square lattice \cite{Johnston:Ap,Dagotto:Rmp,Dai:Rmp,Si:NRM}, where each Fe atom is caged in a tetrahedral structure.
The magnetism of the non-superconducting parent state is primarily given by the collinear stripe-like order, namely the C-type antiferromagnetic (C-AFM) order, although some exceptions exist such as in FeTe, FeSe, KFe$_2$Se$_2$,
and K$_2$Fe$_4$Se$_5$ \cite{Bishop:prl,Li:Prb09,Moon:Prl,Ma:Prl,rong:prl15,Li:Np,Lw:Prb,Zhang:rrl,Bao:Cpl,Ye:Prl,Yin:Prl}.
The structural and magnetic similarities of all these iron-based superconductors suggest common physical mechanisms leading to their magnetic and superconducting properties, despite the different ratios of the atomic elements involved.

\begin{figure}
\centering
\includegraphics[width=0.48\textwidth]{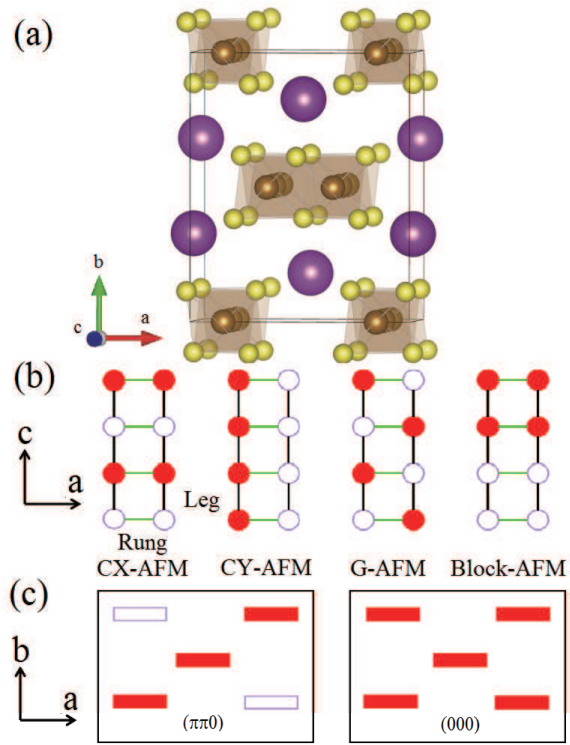}
\caption{(a) Schematic crystal structure of $A$Fe$_2$$X$$_3$.  Purple: $A$, e.g. Ba or K; Yellow: $X$, e.g. S or Se; Brown: Fe.
(b) Sketch of the possible spin configurations that could be stabilized in two-leg iron ladders. Spin up and spin down are distinguished
by colors. (c) Schematic of the three dimensional magnetic ($\pi$, $\pi$, $0$) and ($0$, $0$, $0$)  configurations between ladders.}
\label{Fig1}
\end{figure}

However, recently iron chalcogenides with a different structure, the so-called 123-type $A$Fe$_2X_3$ ($A$=K, Cs, Rb, or Ba; and $X$=S, Se, or Te), have drawn considerable attention \cite{Maziopa:Jpcm,Saparov:Prb,Caron:Prb,Caron:Prb12,Lei:Prb,Arita:prb,Wang:prb16,Patel:prb}. These materials display unique quasi-one-dimensional two-leg ladder iron structures (see Fig.~\ref{Fig1}(a)) that are clearly qualitatively distinct from the other extensively studied iron pnictides/chalcogenides with iron layers. At least from the perspective of the electronic structure, the frequently mentioned Fermi surface nesting effect involving two pocket cylindrical Fermi surfaces (corresponding to the quasi-two-dimensional structure) in several iron pnictides/chalcogenides can not be relevant in these ladder systems \cite{Lumsden:Jpcm,Stewart:Rmp,Johnston:Ap,Dagotto:Rmp,Dai:Rmp}. Thus, the reduced dimensionality of the iron network makes $A$Fe$_2X_3$ a physically fascinating material that deserves further experimental and theoretical scrutiny. In fact, the two-leg ladders made of Fe atoms remind us of the previously studied two-leg ladders superconducting cuprates \cite{Dagotto:Rmp94,Dagotto:Rpp}. In the past, the study of cuprate ladders much illuminated the physics of Cu oxides, primarily because theoretical calculations involving model Hamiltonian can be carried out with good accuracy in chains and ladders and, thus, accurate theory-experiment comparisons can be done. A similar important impact of iron ladders on the field of iron superconductors is now to be expected.

In this family of ladder materials, BaFe$_2$Se$_3$ was the first compound reported to be superconducting at approximately $11$ K \cite{Maziopa:Jpcm}. However, others experiments claimed the material to be a semiconductor with the observed superconductivity probably induced by impurities \cite{Caron:Prb}. Even under these circumstances, one of our previous studies predicted an interesting property, namely that its block-type antiferromagnetic (Block-AFM) state is multiferroic due to magnetostriction effects \cite{Dong:PRL14,Dong:Ap}. The theoretically predicted polar structure has indeed been verified by a subsequent neutron study \cite{Lovesey:PS}. In addition, doping of K in the Ba site changes the ground state to the so-called CX-type antiferromagnetism (CX-AFM) (see Fig. 1(b)) \cite{Caron:Prb12}. Such CX-AFM order was also predicted to be the ground state of BaFe$_2$S$_3$, which was later also confirmed using neutron techniques \cite{Takahashi:Nm}. Under ambient conditions, BaFe$_2$S$_3$ has a semiconducting ground state with a very small gap about $0.06-0.07$ eV  \cite{Gonen:Cm,Reiff:Jssc}. The most striking recent experimental discovery is that high pressure can drive BaFe$_2$S$_3$ to become superconducting when the hydrostatic pressure exceeds about $11$ GPa \cite{Takahashi:Nm} and its highest $T_{\rm c}$ can reach $24$ K at $13.5$ GPa \cite{Yamauchi:prl15}.

A subsequent density functional theory (DFT) study by Suzuki \textit{et al.} on BaFe$_2$S$_3$ confirmed
the CX-AFM nature of the magnetic state (see also our earlier work \cite{Dong:PRL14}) as well as its semiconducting
behavior under ambient conditions \cite{Suzuki:prb}. However, the pressure induced magnetic to nonmagnetic transition
(as phenomenologicall required by superconductivity) was not obvious from that DFT study, although the calculation indeed
showed a semiconductor-metal transition at about $5$ GPa. In those previous DFT calculations, the magnetism persisted with
only a small suppression upon increasing pressure. This problem might be due to the process followed to allow the relaxation
of the crystal structure under pressure, since for simplicity only the positions of sulfur atoms were optimized in the $x$-$y$ plane
in that previous study, which is sufficient for a qualitative analysis. However, in our study presented below all atomic positions were
simultaneously relaxed, allowing us to be not only qualitative but also quantitative in our analysis.

In the present publication, the magnetic properties, electronic structure, and pressure effects corresponding to BaFe$_2$S$_3$
are revisited using first-principles DFT calculations. Our results can be divided in two classes. First, similar results
as in previous DFT efforts and experiments for the ambient conditions have been obtained and confirmed. Second,
a semiconductor-metal transition accompanying the quenching of magnetism has been observed, in good agreement with experimental
observations. This phase transition is probably of first order because we observe a sudden drop of the iron's magnetic moment
and also anomalies in the crystal structure. More specifically, one of our most important results is that the underlying physics
of this transition lies in the modifications by pressure of the effective electronic density at the iron network of relevance.
Thus, the net effect of increasing pressure is equivalent to doping charge into the iron atoms, a nontrivial effect difficult to deduce from
the existing experimental information. Following these arguments, two additional materials $A$Fe$_2X$$_3$ (KFe$_2$S$_3$
and KFe$_2$Se$_3$) have been also studied theoretically by a similar procedure. The quenching of magnetic moments and metallic
phases have also been obtained both for the cases of KFe$_2$S$_3$ and KFe$_2$Se$_3$, although at higher pressures.
The theoretical observation of these pressure induced transitions suggest possible pressure-induced superconducting
states also in KFe$_2$S$_3$ and KFe$_2$Se$_3$ by analogy with what occurs in BaFe$_2$S$_3$, although we cannot explicitly
prove these predictions due to the limitations of DFT techniques to address superconductivity. We expect that our results
should motivate experimental efforts for their confirmation.

\section{Methods}
The DFT calculations are performed based on the generalized gradient approximation (GGA) with the
projector augmented wave (PAW) potentials, as implemented in the Vienna {\it ab initio} Simulation Package
(VASP) \cite{Kresse:Prb,Kresse:Prb96,Blochl:Prb,Perdew:Prl}. The PBE exchange function has been adopted. The plane-wave cutoff is $500$ eV.
The $k$-point mesh is $4\times$3$\times6$ for the minimum crystalline unit cell, which is accordingly adapted for the various magnetic cells
studied (e.g. $2\times$6$\times2$ for CX-AFM ($\pi$, $\pi$, $0$)). Starting from experimental values, both the lattice constants and atomic
positions are fully relaxed until the force on each atom was below $0.01$ eV/{\AA}.

To study the magnetic properties, various possible (in-ladder) magnetic structures are imposed on the Fe ladders,
as shown in Fig.~\ref{Fig1}(b). Despite the in-ladder correlation, the magnetic correlation between ladders can
also slightly affect the physical properties. Therefore, besides the simplest ($0$, $0$, $0$) order,
the ($\pi$, $\pi$, $0$) order is also studied, as indicated in Fig.~\ref{Fig1}(c).

In addition to the standard DFT calculation, the maximally-localized Wannier functions (MLWFs) have also been employed to fit five Fe's $3d$ bands and the Fermi surface, using the WANNIER90 packages \cite{Mostofi:cpc}.

Several previous DFT studies have found that even the pure GGA (or LSDA) procedures overestimates the local magnetic moments
in iron pnictides and chalcogenides \cite{Hansmann:prl10,Mazin:prb,Mazin:np,Dong:Prl,Zhang:rrl}. Thus, using GGA+$U$ with
a positive $U$ will render this inconsistency even more serious. Thus, in some studies, a negative $U$ correction was adopted
to better describe the 122-type Fe-based materials \cite{Ferber:prb,Yi:prb}. Alternatively, the exchange and correlation
kernels were rescaled by an appropriate factor \cite{Ortenzi:prl}. In the present study, both GGA+$U$ and pure GGA have been tested,
and we found that the later provides a better description of $A$Fe$_2$X$_3$ (regarding its crystalline constants, magnetic moments,
as well as band gaps). For this reason, only the pure GGA results will be presented in the rest of this publication.

\section{Results \& discussion}

\subsection{Magnetism \& electronic structure of BaFe$_2$S$_3$}
Our main DFT results with regards to the magnetic and electronic properties of the BaFe$_2$S$_3$ ladder can be summarized in four statements as described below:

(1) Without the external pressure, both the lattice constants and atomic positions were fully optimized in the
presence of magnetism. In the pure GGA calculation, the CX-AFM ($\pi$, $\pi$, $0$)
state has the lowest energy among all candidate configurations investigated, in agreement with experiments.
The CX-AFM ($0$, $0$, $0$) is only slightly higher in energy by $6.3$ meV/Fe, which is
reasonable considering the similarity between these two CX-AFM configurations.

(2) Even the pure GGA result gives a magnetic moment ($2.08$ $\mu_{\rm B}$/Fe) that is slightly higher than
the experimental one ($\sim1.2$ $\mu_{\rm B}$/Fe at low temperature, as measured by neutron scattering \cite{Takahashi:Nm}).
This overestimated local moment is quite common in DFT calculations of iron-based superconductors \cite{Hansmann:prl10,Mazin:prb,Mazin:np},
which may be due to the coexistence of localized Fe spins and itinerant electrons \cite{Tam:prl}. Note that this calculated
value also depends on how large the Wigner-Seitz radius of the iron atom is set to be. Thus, the inconsistency described above
may partially originate from the methodological difference between the VASP procedure and neutron experiments.

(3) For the experimentally relevant CX-AFM ($\pi$, $\pi$, $0$) state, the pure GGA calculation gives a small gap of $0.088$ eV, which agrees
with the experimental value ($0.06-0.07$ meV) \cite{Gonen:Cm,Reiff:Jssc} and previous DFT results \cite{Suzuki:prb}.

(4) According to the calculated density of states (DOS) (not shown), the bands near the Fermi levels are
highly hybridized between Fe's $3d$ orbitals and S's $3p$ orbitals. Such hybridization is quite common in iron pnictides/chalcogenides.

\subsection{Pressure induced transition in BaFe$_2$S$_3$}

Although previous DFT calculations correctly reproduced the magnetic ground state and electronic structure of BaFe$_2$S$_3$,
the pressure induced magnetic-nonmagnetic transition (presumably also associated to superconductivity, although beyond the DFT scope)
was not reproduced \cite{Suzuki:prb}, as explained before. Below, these pressure effects will be revisited using DFT.

By increasing the hydrostatic pressure in the calculation, our energies and Fe's magnetic moments are summarized
in Fig.~\ref{Fig3}(a-b). For all magnetic orders, the local moments decrease with pressure but the CX-AFM ($\pi$, $\pi$, $0$)
state always has the lowest energy. When the pressure reaches $10.8$ GPa (very close to the experimental critical
pressure \cite{Takahashi:Nm,Yamauchi:prl15}),  the {\it static values} of the local moments drop to zero. Meanwhile,
the system becomes metallic. Such nonmagnetic metallic phase provides the conditions for superconductivity if there
are still antiferromagnetic fluctuations present (although short-range magnetic quantum fluctuations also can not be
captured in the DFT calculation). In this sense, without fine tuning parameters, our calculation correctly reproduces
the antiferromagnetic-to-nonmagnetic transition. Furthermore, we found that the quenching of magnetic moments occurs
as in a first-order transition, with a sudden drop from $0.83$ $\mu_{\rm B}$/Fe to zero at $10.8$ GPa (inset of Fig.~\ref{Fig3}(b)).

Figure~\ref{Fig3}(c) shows the compressibility, i.e. the lattice constants normalized to the ambient ones.
The $b$-axis lattice is the softest while the $c$-axis lattice is the hardest, as in experiments \cite{Takahashi:Nm}.
Such anisotropic compressibility can be intuitively understood considering the loose space between ladders and the
compact bonds along ladders. The first-order character of the transition at $10.8$ GPa can also be observed in the lattice
structure (especially for the lattice constant along the $b$-axis), as emphasized in the inset of Fig.~\ref{Fig3}(c).
This first-order characteristic was not reported previously and could be verified in future experiments.

\begin{figure}
\centering
\includegraphics[width=0.44\textwidth]{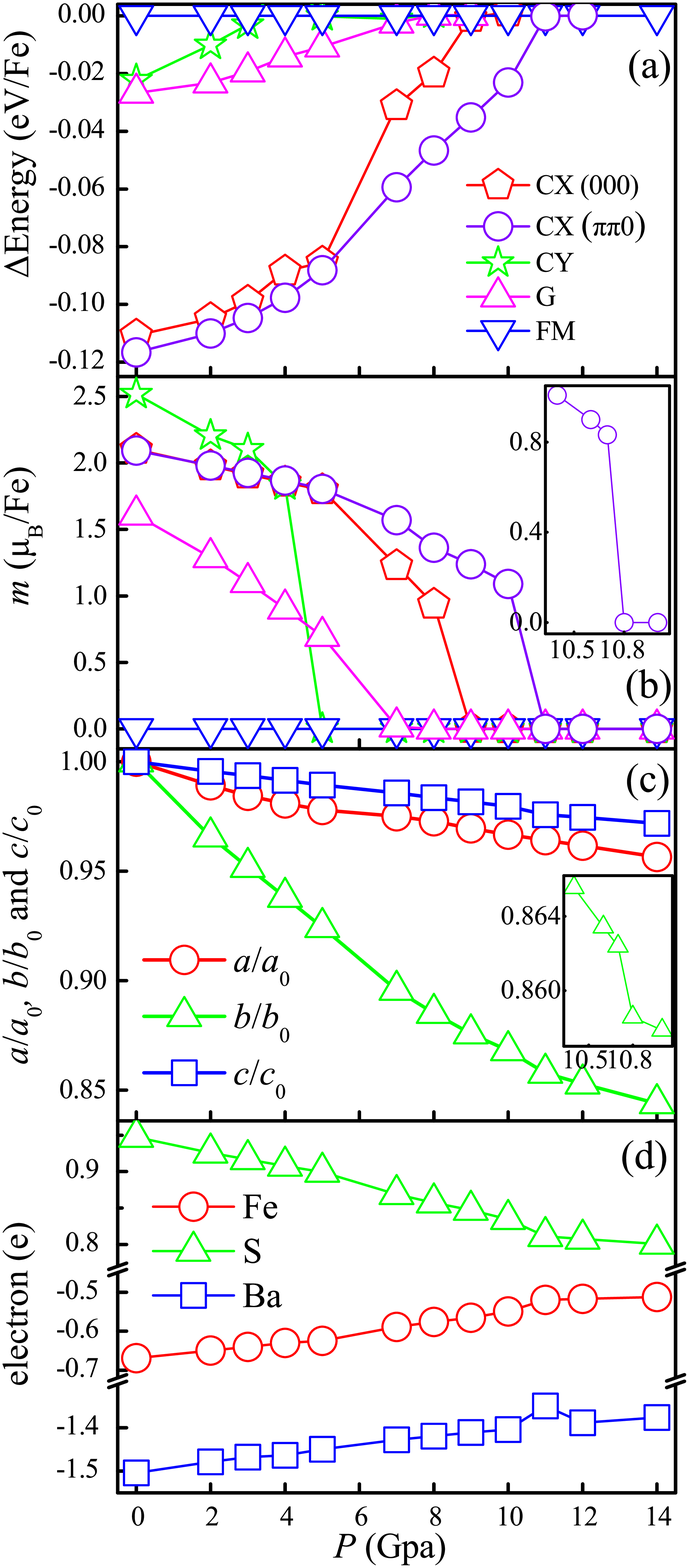}
\caption{DFT results corresponding to BaFe$_2$S$_3$, as a function of pressure. (a) Energy difference (per Fe) of various magnetic orders with respect to the ferromagnetic (FM) state. (b) Local magnetic moments of Fe. Inset: a magnified view near the transition. (c) Lattice constants normalized to the original ones for the CX-AFM ($\pi$, $\pi$, $0$) state. Inset: a magnified view near the transition. (d) Bader charge analysis.}
\label{Fig2}
\end{figure}

According to the Bader charge analysis (Fig.~\ref{Fig2}(d))~\cite{Bader:book,Tang:Jpcm,Henkelman:Cms},
there is a significant charge transfer ($\sim0.15$ electron) from S to Fe by increasing pressure from
$0$ GPa to $12$ GPa. This tendency is equivalent to the effects of electron doping by, for example, chemical substitution,
which is the standard procedure to generate superconductivity from a magnetic parent compound. In this sense,
it is reasonable to suspect that the superconductivity observed in BaFe$_2$S$_3$ probably is induced by electron doping,
in analogy with the superconductivity triggered by F doping in LaFeAsO. This is also quite similar to what happened in
previous investigations of the Cu-ladders, where a transfer of charge from chains to ladders triggers
superconductivity~\cite{Dagotto:sci96}. In summary, our calculations suggest that the superconductivity
of BaFe$_2$S$_3$ is probably caused by self-doping of electrons into the iron network. This effect can occur
in addition to the previously proposed scenario based on the broadening of the electronic bandwidth by
pressure~\cite{Takahashi:Nm}. Only further experimental work can clarify which of the two tendencies is more dominant to generate superconductivity.

A careful analysis of the DOS  at the Fermi level just before
the critical pressure finds a sharp peak, i.e. a van Hove singularity, which suddenly drops
to become a valley around $11$ GPa, as shown in Fig.~\ref{Fig3}(a).
The first-order magnetic-nonmagnetic transition leads to the sudden disappearance of this van Hove singularity
since electronic bands are seriously reconstructed at the critical pressure. The Fermi surface at $11$ GPa
is shown in Fig.~\ref{Fig3}(b), with four bands crossing the Fermi level. Two of them are nesting around
the $\Gamma$ point while the other two are isolated.

\begin{figure}
\centering
\includegraphics[width=0.44\textwidth]{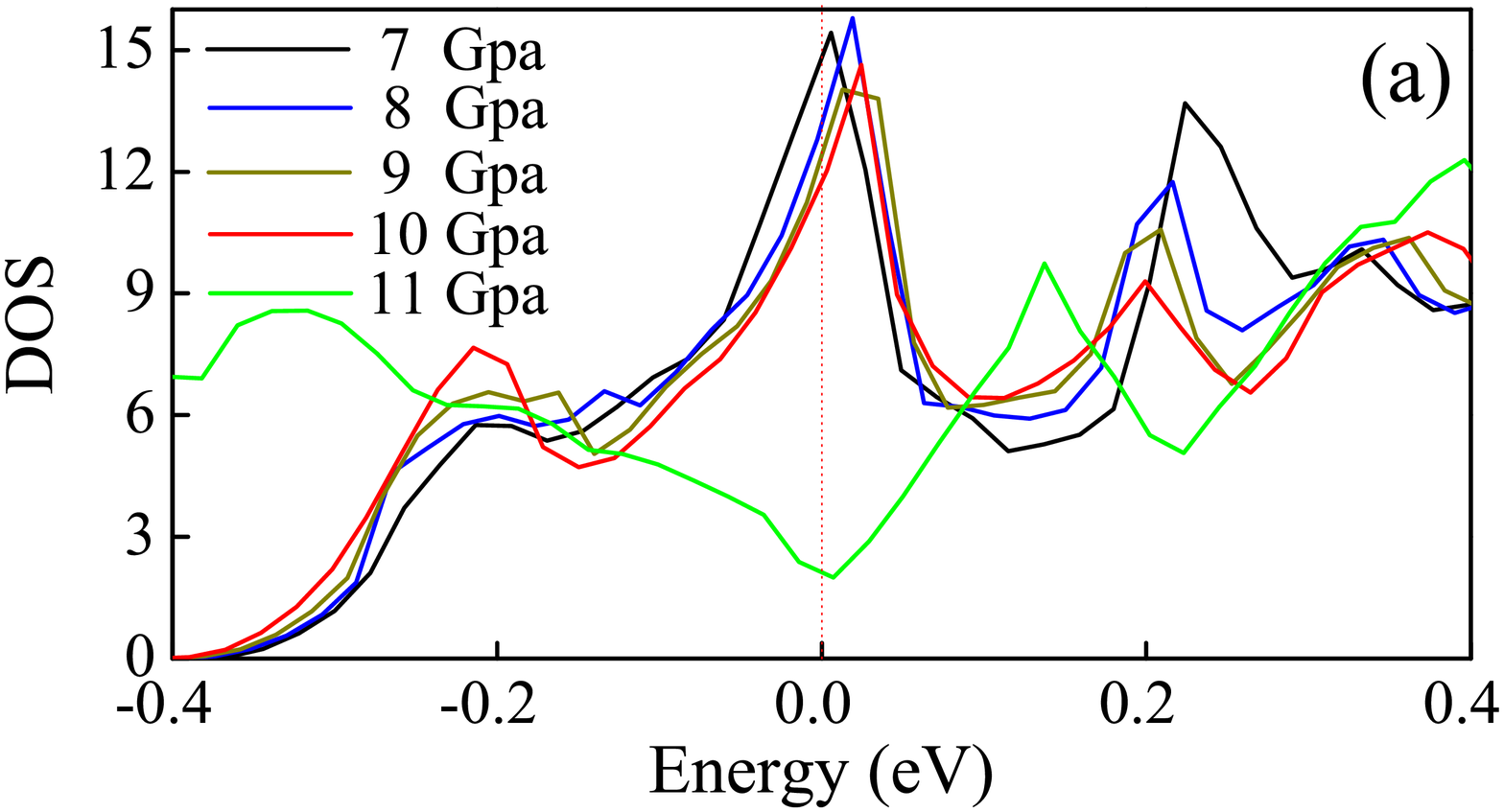}
\includegraphics[width=0.48\textwidth]{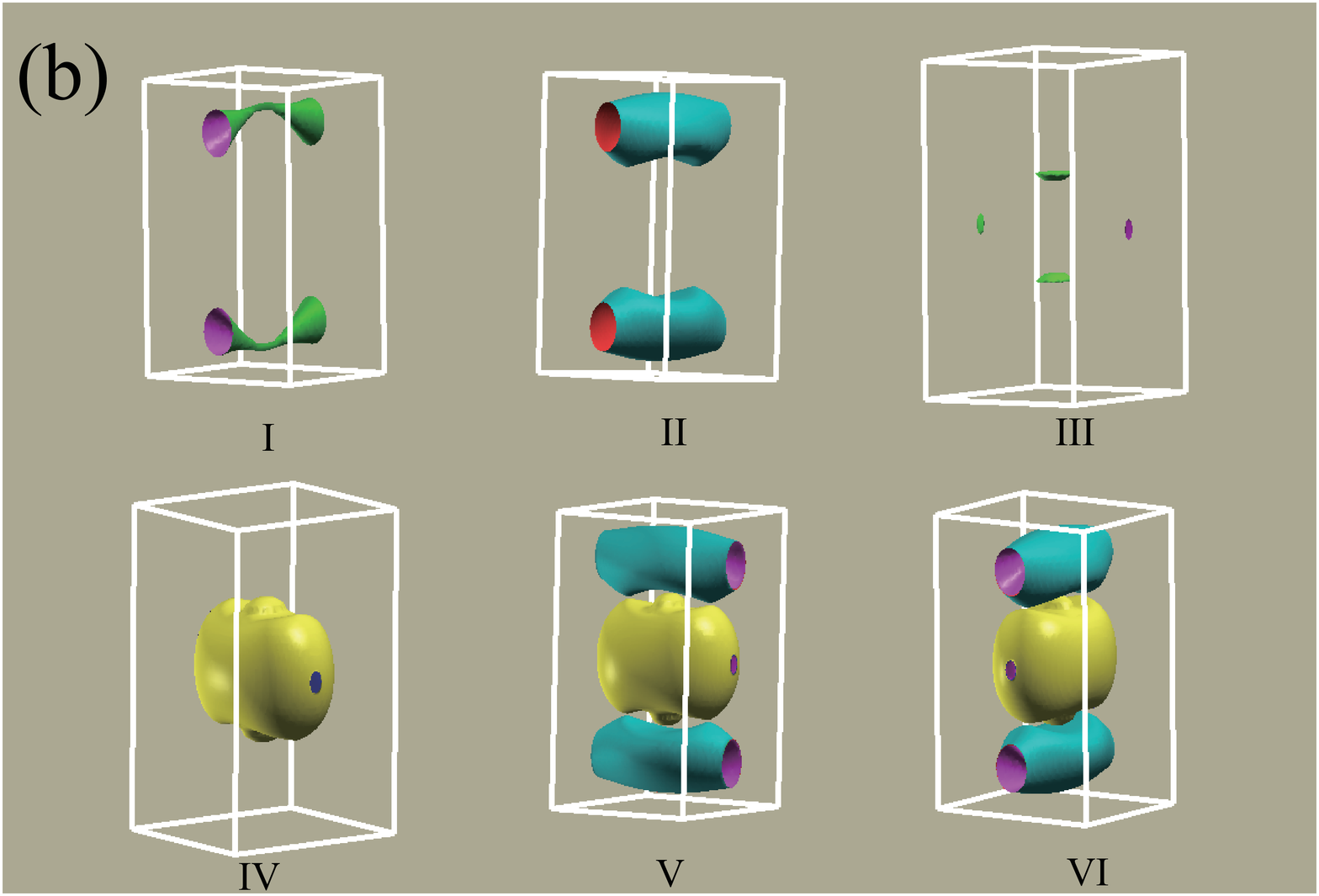}
\caption{(a) Evolution of the DOS near the Fermi level with increasing pressure. Below the critical pressure,
a van Hove singularity appears at the Fermi level, but it suddenly disappears at the critical pressure. Only
the spin-up channel is shown because the spin-down channel is exactly identical, since the system is either
antiferromagnetic or nonmagnetic. (b) Fermi surfaces at $11$ GPa. The first four (I-IV) panels are individual
Fermi surfaces. The last two (V-VI) are the total Fermi surfaces viewed from different orientations.}
\label{Fig3}
\end{figure}

\subsection{Pressure induced transition in KFe$_2$S$_3$}

Until now, BaFe$_2$S$_3$ is the only experimentally confirmed superconductor in the 123-type series of iron ladders. As a consequence, it is interesting to investigate whether there are other 123-type iron ladders that can also potentially become superconductors. KFe$_2$S$_3$ is a sister member of BaFe$_2$S$_3$, where the nominal valence of Fe is higher by $+0.5$. Experimentally, KFe$_2$S$_3$ has been synthesized \cite{Mitchell:Jssc} but their detailed physical properties have not been reported, particularly under high pressure. Structurally, this K-based 123 ladder is similar to BaFe$_2$S$_3$, with the same $Cmcm$ group.

Here, DFT calculations have been performed also on KFe$_2$S$_3$. In this study we have found that
the magnetic ground state is also of the CX-AFM type. Moreover, the local magnetic moment is about $2.3$ $\mu_{\rm B}$/Fe
in the pure GGA calculations. Compared to BaFe$_2$S$_3$, the band gap of KFe$_2$S$_3$ ($\sim0.51$~eV in pure GGA calculations)
is slightly larger. According to the atomically-resolved DOS (not shown), the states near the Fermi level are also primarily
contributed by the Fe's $3d$  orbitals, which are also highly hybridized with the S's $3p$ orbitals.

According to the Bader charge analysis, the electronic densities at the Fe and S sites in KFe$_2$S$_3$ are lower by $0.09$ and $0.16$ electrons
than those in BaFe$_2$S$_3$, respectively. Therefore, the replacement of Ba$^{2+}$ by K$^{1+}$ does not really dope the iron sites by the nominal
amount of $0.5$ holes, but those holes mostly go to the S's sites due to the partially covalent Fe-S bonds.

\begin{figure}
\centering
\includegraphics[width=0.48\textwidth]{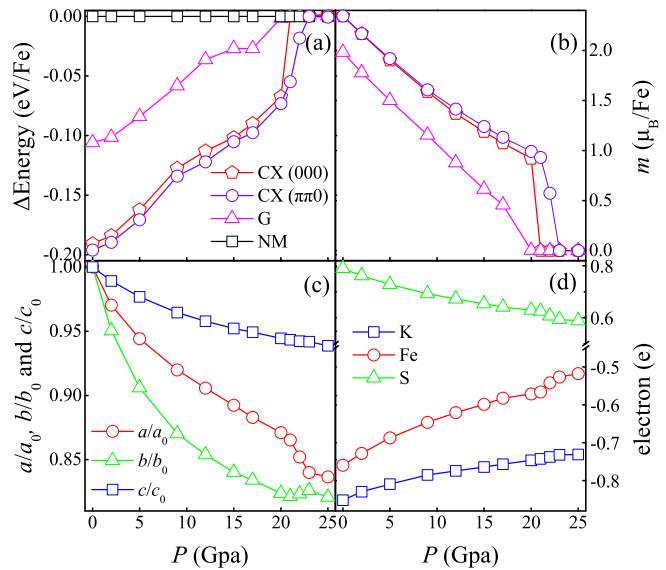}
\caption{DFT results of KFe$_2$S$_3$ as a function of pressure. (a) Energies (per Fe) of various magnetic orders. (b) Local magnetic moments of Fe. (c) Lattice constants normalized to the original ones for the CX state. (d) Bader charge analysis.}
\label{Fig4}
\end{figure}

Since the electronic density of Fe is slightly lower in KFe$_2$S$_3$, the critical pressure should
be higher according to the Bader charge analysis. To verify this expectation, the calculated crystal constants and
magnetism (with pure GGA) are presented in Fig.~\ref{Fig4} as a function of pressure.
As shown in Fig.~\ref{Fig4}(a), the CX-AFM ($\pi$, $\pi$, $0$) state is always the ground state if it is magnetically ordered.
The suppression of the magnetic moment by pressure is shown in Fig.~\ref{Fig4}(b). The first-order character of the transition
is similar to that observed in BaFe$_2$S$_3$. The critical pressure ($\sim23$ GPa) is indeed larger, as expected. The band gap
of the CX-AFM phase reduces to zero at $17$ GPa, inducing a semiconductor to metal transition. The metallic nonmagnetic phase
above $23$ GPa may be superconducting, according to the previous experience with BaFe$_2$S$_3$. Of course, this reasoning is merely
by analogy between similar materials because DFT cannot address superconductivity directly.
The lattice constants under pressure, normalized to their ambient values, are shown in Fig.~\ref{Fig4}(c); they are also very similar
to those reported for BaFe$_2$S$_3$. The Bader charge analysis applied to KFe$_2$S$_3$ under pressure leads to the same behavior as
in BaFe$_2$S$_3$, namely pressure enhances the local electronic density at the Fe sites, as shown in Fig.~\ref{Fig4}(d). There is also
significant charge transfer ($\sim0.24$ electron per Fe) from S to Fe by increasing pressure from $0$~GPa to $25$~GPa. Interestingly,
the Bader charge densities at the critical pressures for magnetic quenching are almost identical (error bar $\delta<0.005e$ in our
calculations) for KFe$_2$S$_3$ and BaFe$_2$S$_3$, suggesting a similar physical mechanism for both compounds.

To summarize this subsection, our calculations here predict that KFe$_2$S$_3$ should be similar to BaFe$_2$S$_3$, regarding its
magnetic ground state as well as its behavior upon pressure. Thus, superconductivity is possible under higher pressure.
By increasing pressure, the transfer of electrons from S to Fe occurs, namely a self-doping process takes place that eventually
could lead to superconductivity as in the canonical layered iron superconductors.

\subsection{Pressure induced transition in KFe$_2$Se$_3$ }

Since reducing the electronic density at the Fe atoms is a disadvantage to suppress magnetism with increasing pressure,
as demonstrated in KFe$_2$S$_3$ where a higher pressure than for BaFe$_2$S$_3$ was needed to induce the metallic phase,
it is natural to expect the opposite tendency in other $A$Fe$_2X_3$ compounds that naturally have higher electronic density at the Fe atoms.

According to our Bader charge analysis, at ambient conditions the electronic density at the Fe atoms
in KFe$_2$Se$_3$ (experimentally confirmed to display the CX-AFM order~\cite{Caron:Prb12}) is higher than
in the case of KFe$_2$S$_3$ by $0.12$ electrons, and even higher than that in BaFe$_2$S$_3$ by $0.03$ electrons,
due to the weak electronegativity of Se. Then, a natural speculation arises: could it be that by increasing pressure
KFe$_2$Se$_3$ is closer to metallicity, and thus perhaps also superconductivity, than BaFe$_2$S$_3$ is?

\begin{figure}
\centering
\includegraphics[width=0.48\textwidth]{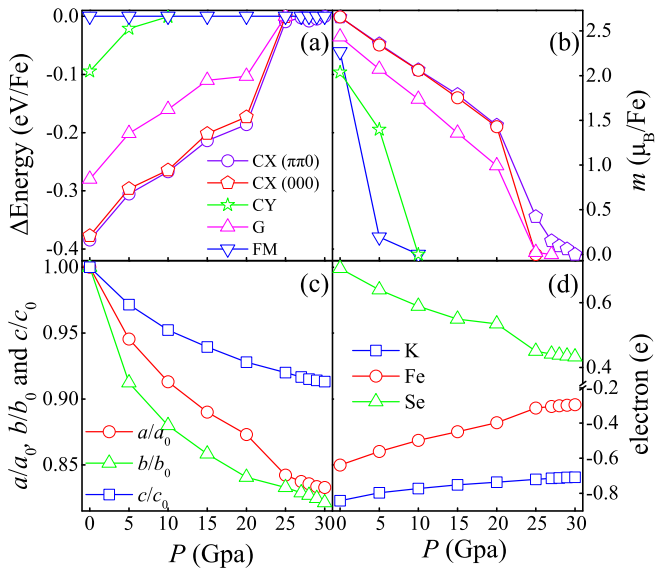}
\caption{DFT results of KFe$_2$Se$_3$ as a function of pressure. (a) Energies (per Fe) of various magnetic orders. (b) Local magnetic
moments of Fe. (c) Lattice constants normalized to the original ones for the CX state. (d) Bader charge analysis.}
\label{Fig5}
\end{figure}

Despite the experimental studies by Caron \textit{et al.}~\cite{Caron:Prb12}, DFT calculations on KFe$_2$Se$_3$
have not been performed to our knowledge. To remedy this problem, here the pure GGA calculation has been carried
out for KFe$_2$Se$_3$. At ambient conditions, the CX-AFM state is indeed the ground state (Fig.~\ref{Fig5}(a)).
The local moment of Fe is large, reaching the value $2.65$ $\mu_{\rm B}$/Fe (slightly higher than the experimental
one $2.1$ $\mu_{\rm B}$/Fe~\cite{Caron:Prb12}) at ambient conditions (Fig.~\ref{Fig5}(b)), which is a negative signal for metallicity,
and thus for potential superconductivity.

Upon pressure, the quenching of the magnetic moment and semiconductor-metal transition indeed occurs. The gap of KFe$_2$Se$_3$ is about $0.56$ eV at ambient condition, which is gradually closed by increasing pressure to $25$~GPa. The required critical pressure for magnetic quenching reaches $29$~GPa, even higher than that of KFe$_2$S$_3$. And a different feature is that this magnetic phase transition seems to be more gradual, probably of second order or weak first order, rather than occurring by a sudden jump as observed in BaFe$_2$S$_3$ and KFe$_2$S$_3$. The Bader charge analysis is shown in Fig.~\ref{Fig5}(d) as a function of pressure. Furthermore, for the higher magnetic moment in KFe$_2$Se$_3$, the critical pressure should be higher to suppress the magnetism. Then, in this case additional charge transfer from Se to Fe may be required to suppress the magnetism.

In summary of this subsection, our DFT calculations have confirmed the CX-AFM magnetic ground state for KFe$_2$Se$_3$. Although KFe$_2$Se$_3$ owns a relative high electronic density at the Fe atoms, its large gap and large moment make it even more difficult to induce a nonmagnetic metallic phase upon pressure. And this magnetic-nonmagnetic transition may be of the second order, or weak first order. These different features may arise from the Se atoms, which are larger in size and weaker in their electronegativity.

\section{Conclusion}
In this work, the magnetic and electronic properties of BaFe$_2$S$_3$, KFe$_2$S$_3$, and KFe$_2$Se$_3$ have been
analyzed using first-principles calculations. The CX-AFM magnetic order is confirmed to be the common
magnetic ground state for all these materials. The pressure-driven semiconductor to metal transition, as well as
the antiferromagnetic-to-nonmagnetic transition, has been properly reproduced. Although the DFT technique
can not directly address a superconducting state, our study can provide helpful information to understand
the superconducting transition of BaFe$_2$S$_3$ at high pressure ($11$~GPa), which is predicted to be
a first-order transition. Our main conclusion is that the electron transfer from S to Fe, i.e. a self-doping
process, may play a key role to tune the magnetism in BaFe$_2$S$_3$ and eventually induce metallicity and potentially
superconductivity.

A similar first-order transition has also been predicted for KFe$_2$S$_3$, although the required critical pressure
is higher (about $23$~GPa). By contrast, although the magnetism can also be quenched in KFe$_2$Se$_3$, the required
pressure (about $29$~GPa) is even higher and the transition seems to be of the second order, or weak first order.
Further experiments are encouraged to verify our predictions as well as the possible existence of metallicity,
and perhaps superconductivity, in KFe$_2$S$_3$ and KFe$_2$Se$_3$.

\acknowledgments{We acknowledge valuable discussions with Rong Yu, Hongyan Lu, and Songxue Chi. This work was primarily
supported by the National Natural Science Foundation of China (Grant No. 11674055). E.D. was supported by the U.S. Department of
Energy, Office of Basic Energy Sciences, Materials Sciences and Engineering Division. Most calculations
were performed on Tianhe-2 at the National Supercomputer Center in Guangzhou (NSCC-GZ).}

\bibliographystyle{apsrev4-1}
\bibliography{ref3}

\end{document}